\documentclass[12pt]{article}
\usepackage{epsfig}
\begin{document}
\begin{center}
{\Large \bf Invariant spin coherent states and the theory
of quantum antiferromagnet in a paramagnetic phase}

{\large V. I. Belinicher$^{*\dagger}$, and J. da Providencia$^{*}$}
\end{center}

\noindent
$^{*}$ University of Coimbra, 3000, Coimbra, Portugal \\
$^{\dagger}$Institute of Semiconductor Physics, 630090, Novosibirsk,
Russia

\begin{center}
{\Large \bf Abstract}
\end{center}

 The consistent theory of the Heisenberg quantum
antiferromagnet in the disordered phase with short range
antiferromagnetic order was developed on the basis of the path integral for the spin
coherent states. We have presented the Lagrangian of the theory in
a form which is explicitly invariant under rotations and have found natural
variables in the term of which one can construct a natural perturbation
theory. The short wave spin  fluctuations are similar to the spin wave
theory ones, and the long wave spin fluctuations are governed by the nonlinear
sigma model.
We have also demonstrated that the short wave spin fluctuations
have to be considered accurately in the framework of the discrete
version in time of the path integral.
In the framework of our approach we have obtained the response function for the
spin fluctuations for the whole region of the frequency $\omega$ and the wave
vector ${\bf k}$ and have calculated the free energy of the system.

\vskip 0.5cm
\noindent
Pacs: 75.50.Ee,74.20.Mn \\ \ \\

The theory of the two--dimensional Heisenberg antiferromagnet (AF) has
attracted great interest during the last years in connection
with the problem of AF fluctuations in
copper oxides \cite{Man1}, \cite{Cha1}, \cite{Sac1}.
The approach of these papers was based on the sigma model,
which describes the long wave fluctuations of the Heisenberg AF
in the paramagnetic phase with a short range antiferromagnetic order.
The sigma model is the continuum model for the unit vector
${\bf n}(t,{\bf r}), \ {\bf n}^2 =1$ in the 1 +
2 time and space dimensions \cite{Pol1,Zin1}. As a long wave theory,
the sigma model can make a lot of physical predictions such as the
structure of the long wave fluctuations and the magnitude of the
correlation length \cite{Cha1,Sac1,Has1}. But up to now a consistent
theory of the spin fluctuations for the quantum AF (QAF) with short
range AF order was absent. This is just the topic of this paper.

Our approach to the description of the QAF is based on the
functional integral for the generalized partition function (GPF) in terms
of spin coherent states. We introduce the concept of
invariant spin
coherent states and on this basis we formulate the theory.

We define the invariant spin coherent states (SCS) with the help of
relation:
\begin{eqnarray} \label{1}
|{\bf n};{\bf m}>=\exp(-i\varphi\hat{S}_z)\exp(-i\theta\hat{S}_y).
\exp(-i\psi\hat{S}_z)|ss>.
\end{eqnarray}
Here, the state $|ss>$ is the state of spin $s$ with the maximal spin
projection $s$. The unit vectors ${\bf n}$ and ${\bf m}$ are
orthogonal: ${\bf n}^2=1,\ {\bf m}^2=1, \ {\bf n}\cdot{\bf m}=0$.
$\theta, \varphi$ are the Euler angles of the unit vector
${\bf n}=(\cos\varphi\sin\theta,\sin\varphi\sin\theta,\cos\theta)$.
The dependence on the vector ${\bf m}$ is included in the angle $\psi$
only, which, in fact determines only the phase factor in the SCS (\ref{1}).  We
can choose the angle $\psi$ in some special manner which distiguishes this
definition from the standard one \cite{Kla1}: $\tan\psi =-k_z/m_z$, where
the vector ${\bf k}=[{\bf n}\times{\bf m}]$. This choice has a clear
geometrical interpretation. The transformation (\ref{1}) rotates the
reference coherent 
state which is characterized by the vectors
${\bf n}_0=(0,0,1)$ and ${\bf m}_0=(1,0,0),$ into the SCS (\ref{1}).
>From this geometric interpretation it is obvious that upon changing the
vector ${\bf n}$ into the vector ${\bf n}'$ with the help of some rotation
$\hat{a},$ we have,
\begin{eqnarray} \label{2}
|\hat{a}{\bf n};\hat{a}{\bf m}>=\hat{U}(\hat{a})|{\bf n};{\bf m}>
\end{eqnarray}
without the phase factor which was introduced and discussed by Perelomov
\cite{Per1}.
It seems that the vector ${\bf m}$ is an artificial one. However, for the
problem of the QAF it has some real meaning.

We consider the spin system which is described by the Heisenberg Hamiltonian:
\begin{eqnarray} \label{3}
\hat{H}_{Hei} = \frac{J}{2}\sum_{l,l'=<l>}\hat{\bf S}_{l} \cdot
\hat{\bf S}_{l'}, \ \ \ \hat{\bf S}_{l}\cdot\hat{\bf S}_{l} = s(s+1),
\end{eqnarray}
where $\hat{\bf S}_{l}$ are the spin operators; the index $l$ runs over
a two--dimensional square lattice; the index $l'$ runs over
the nearest neighbors of the site $l$;
$J>0$ is the exchange constant which, since it is positive,
corresponds to the AF spin interaction; and $s$ is the magnitude of spin.
 The most efficient method of dealing with a spin system is based on the
representation of the GPF $Z$ or the generating functional of the
spin Green functions in the form of a functional
integral over spin coherent states
\begin{eqnarray}  \label{4}
&&Z = Tr\left[\exp\left(-\beta\hat{H}\right)\right], \ \ \
\beta = 1/T,
\\  \label{5}
&&Z = \int_{-\infty}^{\infty} \cdots
\int_{-\infty}^{\infty}
D\mu({\bf n}_a,{\bf n}_b) \exp(A({\bf n}_a,{\bf n}_b)),
\\  \label{6}
&&D\mu({\bf n}_a,{\bf n}_b)=\prod_{p=a,b;\tau,l} \frac{2s+1}{2\pi}
\delta({\bf n}_p^2(\tau,l)-1)d{\bf n}_p(\tau,l)
\end{eqnarray}
where $T$ is the temperature, $\tau$ is the imaginary time, and $A({\bf
n})$ is the action of the system. In the continuum approximation,
which is valid in the leading order in $1/2s$ the expression of the action
$A({\bf n})$ is simplified
\begin{eqnarray} \label{7}
&&A({\bf n}_a,{\bf n}_b) = -\int_{0}^{\beta}\sum_{l}{\cal L}_{tot}(\tau,l)d\tau, \ \ \
{\cal L}_{tot}(\tau,l)={\cal L}_{kin}(\tau,l)+{\cal H}(\tau,l),
\\ \label{8}
&&{\cal L}_{kin}(\tau,l) = \sum_{p=a,b}
<{\bf n}_p(\tau,l);{\bf m}_p(\tau,l)|\frac{\partial}{\partial\tau}
|{\bf n}_p(\tau,l);{\bf m}_p(\tau,l)>,
\\ \nonumber
&&{\cal H}(\tau,l) = Js^2\sum_{l'=<l>}{\bf n}_a(\tau,l)
\cdot{\bf n}_b(\tau,l').
\end{eqnarray}
The idea of the short range AF order was used in Eqs. (\ref{5}-\ref{8}),
and we split our square lattice into two AF sublattices
$a$ and $b$.
For the kinetic part of the action ${\cal L}_{kin}$ (which is highly
nonlinear) we use the concept of invariant coherent state parametrized by
arbitrary vectors ${\bf m}_{a,b}$.

In our case we can define these vectors in the following manner
${\bf m}_{a,b}$:
${\bf m}_{a}=R_{ab}[{\bf n}_{b}-({\bf n}_{b}\cdot{\bf n}_{a}){\bf n}_{a}]$,
${\bf m}_{b}=R_{ab}[{\bf n}_{a}-({\bf n}_{a}\cdot{\bf n}_{b}){\bf n}_{b}]$,
and $R_{ab}=[1-({\bf n}_{a}\cdot{\bf n}_{b})^2]^{-1/2}$, and the invariant
coherent states have a clear meaning. Substituting these expressions for
${\bf m}_{a,b}$ into Eq. (\ref{8}) for ${\cal L}_{kin}$ we have also the
invariant expression ${\cal L}_{kin}$
\begin{eqnarray} \label{9}
{\cal L}_{kin}=\frac{is}{1-{\bf n}_{a\tau l}\cdot{\bf n}_{b\tau l}}
(\dot{\bf n}_{a\tau l}-\dot{\bf n}_{b\tau l})\cdot
[{\bf n}_{a\tau l}\times{\bf n}_{b\tau l}].
\end{eqnarray}
Now we can introduce new more convenient variables
$\mbox{\boldmath{$\Omega$}}(\tau,l)$ and
${\bf M}(\tau,l)$ which realize the stereographic
mapping of a sphere:
\begin{eqnarray} \label{10}
{\bf n}_{a,b} = \frac{\pm \mbox{\boldmath{$\Omega $}}
\left(1-{\bf M}^2/4\right)-[\mbox{\boldmath{$\Omega $}}\times
{\bf M}]}{1+{\bf M}^2/4}, \ \ \ \mbox{\boldmath{$\Omega $}}^2=1, \ \ \
\mbox{\boldmath{$\Omega $}}\cdot{\bf M}=0.
\end{eqnarray}

In terms of these variables the total Lagrangian ${\cal L}_{\Omega M}=
{\cal L}_{kin}+{\cal H}$ has the final form
\begin{eqnarray} \label{11}
&{\cal L}_{kin}=&\frac{2is\dot{\mbox{\boldmath{$\Omega$}}}\cdot{\bf M}}
{1+{\bf M}^2/4}, \ \ \ {\cal H}=
Js^2\sum_{l'=<l>}\{\mbox{\boldmath{$\Omega$}}\cdot
\mbox{\boldmath{$\Omega$}}'[(1-{\bf M}^2/4)(1-{\bf M}'^2/4)
\\ \nonumber
&&-{\bf M}\cdot{\bf M}']+
\mbox{\boldmath{$\Omega$}}\cdot{\bf M}' \
\mbox{\boldmath{$\Omega$}}'\cdot{\bf M}\}(1+{\bf M}^2/4)^{-1}
(1+{\bf M}'^2/4)^{-1},
\end{eqnarray}
where $\mbox{\boldmath{$\Omega$}}\equiv\mbox{\boldmath{$\Omega$}}
_{\tau l}, \ \mbox{\boldmath{$\Omega$}}'\equiv\mbox{\boldmath{$\Omega$}}
_{\tau l'}, \ {\bf M}\equiv{\bf M}_{\tau l}, \ {\bf M}'
\equiv{\bf M}_{\tau l'}$.
After this change of variables the measure of integration $D\mu({\bf n})$
(\ref{10}) becomes
\begin{eqnarray} \label{12}
D\mu({\bf n})= \prod_{\tau l}\frac{(2s+1)^2}{2\pi^2}
\frac{1-{\bf M}^2/4}{(1+{\bf M}^2/4)^3}
\delta\left(\mbox{\boldmath{$\Omega$}}^2-1\right)
\delta\left(\mbox{\boldmath{$\Omega$}}\cdot{\bf M}\right)
d\mbox{\boldmath{$\Omega$}}d{\bf M},
\end{eqnarray}
where the product in (\ref{12}) is performed over the AF
(doubled) lattice cells.

The variable $\mbox{\boldmath{$\Omega$}}$ is responsible for the
AF fluctuations and the variable ${\bf M}$ for the
ferromagnetic ones. The ferromagnetic fluctuations are small
according to the
parameter $1/2s$ and therefore one can
expand the Lagrangian ${\cal L}_{\Omega M}$ (\ref{11}) over ${\bf M}$.
The vector of the ferromagnetic fluctuations ${\bf M}$ plays the role (up to
the factor 2s) of the canonical momentum conjugate to the canonical
coordinate $\mbox{\boldmath{$\Omega$}}$.  The term of first order in
${\bf M}$ coincides (after change of variables) with previous results
\cite{Man1,Sac1}.

 From Eq. (\ref{1}) one can easily extract the quadratic part
of the total lagrangian in
the variables $\mbox{\boldmath{$\Omega$}}$ and ${\bf M},$  ${\cal L}_{quad},$
\begin{eqnarray} \label{13}
{\cal L}_{quad} = 2is({\bf M}\cdot\dot{\mbox{\boldmath{$\Omega$}}}) +
Js^2\sum_{l'\in<l>}\left[
\mbox{\boldmath{$\Omega$}}^2-\mbox{\boldmath{$\Omega$}}\cdot
\mbox{\boldmath{$\Omega$}}' + {\bf M}^2 + {\bf M}\cdot{\bf M}'\right],
\end{eqnarray}

 The Lagrangian ${\cal L}_{quad}$ (\ref{13}) is
very simple but the measure $D\mu$ (\ref{12}) is not simple
due to the presence of two delta-- functions. Therefore we cannot
simply perform the Gaussian integration over the fields
$\mbox{\boldmath{$\Omega$}}$ and ${\bf M}$.
To solve this problem we shall use the method of the Lagrange multiplier
$\lambda$ together with the saddle point approximation \cite{Pol1,Zin1} to
eliminate $\delta(\mbox{\boldmath{$\Omega$}}^2-1)$. As a result, we shall
have an additional integration over $\lambda $ with the additional
Lagrangian
\begin{eqnarray} \label{14}
{\cal L}_{\lambda}(\tau,l)=[i\lambda(\tau,l)+\mu_0^2/2{\cal J}]
[\mbox{\boldmath{$\Omega$}}^2(\tau,l)-1],
\end{eqnarray}
where $\mu_0$ is the primary mass of the $\Omega$ field, and
${\cal J}=Jsz$.

To eliminate $\delta(\mbox{\boldmath{$\Omega$}}\cdot{\bf M})$ we shall
use some kind of Faddev--Popov trick \cite{Zin1}.
As a result of this trick: (1) the factor
$\delta\left(\mbox{\boldmath{$\Omega$}}\cdot{\bf M}\right)$ disappears
from the measure (\ref{12}); (2) ${\bf M}\Rightarrow{\bf M}_{tr}={\bf
M}-\mbox{\boldmath{$\Omega $}} (\mbox{\boldmath{$\Omega $}}\cdot{\bf M})$
in the Lagrangian (\ref{11}); (3) an additional contribution to the action
appears, the Lagrangian of which ${\cal L}_{gaug}$ is conveniently
chosen in the form
\begin{eqnarray} \label{15}
{\cal L}_{gaug}=Js^2\sum_{l'\in<l>}\left[
(\mbox{\boldmath{$\Omega$}}\cdot{\bf M})^2 +
(\mbox{\boldmath{$\Omega$}}\cdot{\bf M})
(\mbox{\boldmath{$\Omega$}}'\cdot{\bf M}')\right],
\end{eqnarray}
such choice kills the major dependence on
$\mbox{\boldmath{$\Omega$}}$ in the Lagrangian (\ref{13}) which appears
due to substitution ${\bf M} \Rightarrow {\bf M}_{tr}$; (4)
in the measure of the integration in (\ref{12}) the additional factor
$(\det(\hat{B}_{gaug}))^{1/2}$, where the operator $\hat{B}_{gaug}$ is
just the operator in the quadratic form in the variable
$(\mbox{\boldmath{$\Omega$}}\cdot{\bf M})$ in (\ref{15}).
In this way, the expression (\ref{13}) for
${\cal L}_{quad}$ is valid in the
leading order with respect to $1/2s$.
The final expression for
the total Lagrangian is ${\cal L}_{tot}= {\cal L}_{\Omega M} +
{\cal L}_{gaug}+{\cal L}_{\lambda}$ (\ref{11},\ref{14},\ref{15}).

 Now, from the quadratic part of the total Lagrangian ${\cal L}_{tot}$
one can find the Green functions of the $\Omega$ and $M$ fields
\begin{eqnarray} \label{16}
&&\hat{G}_q{\bf X}^*_q\equiv \left(\begin{array}{cc} G^{\Omega}_q, & G^d_q
\\ G_q^u, & G^M_q \end{array} \right)
\left(\begin{array}{c} \mbox{\boldmath{$\Omega$}}^*_q \\
{\bf M}^*_q\end{array} \right)=
\frac{1}{2sL_q} \left(\begin{array}{cc} Q_{\bf k}, & -\omega \\
\omega, & P'_{\bf k}\end{array} \right),
\\ \nonumber
&&L_q=\omega^2+\omega^2_{0{\bf k}}, \ \ \
\omega^2_{0{\bf k}}=P'_{\bf k}Q_{\bf k}=
(1-\gamma^2_{\bf k}){\cal J}^2+(1+\gamma_{\bf k})\mu_0^2/2.
\\ \nonumber
&&(Q_{\bf k},P_{\bf k}) = {\cal J}(1 \pm \gamma_{\bf k}), \ \ \
\gamma_{\bf k} = (1/2)(\cos(k_xa)+\cos(k_ya)).
\end{eqnarray}

>From Eq. (\ref{16}) one can calculate the parameter of the spin wave
nonlinearity of the theory: $<{\bf M}^2_{tr}>=(1/2s)C_{M}(T)$,
where $C_{M}(T)=0.65075$ for $T \ll {\cal J}$, and
$C_{M}(T)=1.48491T/{\cal J}$ for $T \ge {\cal J}$.

We also have the saddle point condition for the $\lambda$ field
$<\mbox{\boldmath{$\Omega$}}^2>=1$ which is the
most important constraint of the theory which determines its phase state:
\begin{eqnarray} \label{17}
&&1=<\mbox{\boldmath{$\Omega$}}^2>=N\sum_{q}G^{\Omega}_q =
\frac{NT}{2s}\sum_{\omega=2\pi nT}
\sum_{\bf k}\frac{Q_{\bf k}}{\omega^2+\omega^2_{0{\bf k}}},
\\ \label{18}
&&1=<\mbox{\boldmath{$\Omega$}}^2>=\frac{N}{2s}\sum_{\bf k}
\frac{Q_{\bf k}}{2\omega_{0{\bf k}}}(1+2n_{0{\bf k}}),
\end{eqnarray}
where $n_{0{\bf k}}=\left(\exp(\omega_{0{\bf k}}/T)-1\right)^{-1}$
is the Plank function.
 The right hand side of Eq. (\ref{18}) contains
two terms. The first term $Q_{\bf k}/2\omega_{0{\bf k}}$ is responsible
for the quantum fluctuations of the $\Omega$ fields.The second term
$Q_{\bf k}n_{0{\bf k}}/\omega_{0{\bf k}}$ is responsible for the classical
thermal fluctuations of the $\Omega$ fields. The role of these two terms
is quite different. The quantum fluctuations are small according to the
parameter of perturbation theory $1/2s$ and, for the basic approximation,
they can be neglected. The thermal fluctuations can be considered in the
continuum approximation which leads to the well known
\cite{Man1,Cha1,Sac1} zero order expression for $\mu_0$
\begin{eqnarray} \label{19}
\mu_0 = T\exp\left(-\frac{2\pi Js^2}{TN}\right), \ \ \ \xi =
\hbar c_{s}/\mu,
\end{eqnarray}
where $\xi$ is the correlation length.
From Eq. (\ref{19}) the important conclusion is followed: {\it in
the regime of the weak coupling the correlation length} $\xi$ {\it is much
larger than the lattice constant} $a$.

 To close the theory it is helpful to define the polarization operator
$\Pi(q)$ of the $\Omega$ field
\begin{eqnarray} \label{20}
A_{\lambda quad}=-\frac{1}{2}\sum_{q}\lambda^*(q)\Pi(q)\lambda(q),
\end{eqnarray}
and the Green function of the $\lambda$
field is $\Pi(q)^{-1}$. In the lowest approximation $\Pi(q)$ is simply
a loop from two Green function $G^{\Omega}$
\begin{eqnarray} \label{21}
\Pi_0(q)=2NT\sum_{q'}G^{\Omega}(q')G^{\Omega}(q-q').
\end{eqnarray}
The main contribution in $1/2s$
in (\ref{21}) is from the thermal fluctuations even at low temperatures
$T$, because the integral strength of such fluctuations is fixed by the
saddle point condition (\ref{17}) and does not depend on the temperature.
The explicit form for $\Pi_0(q)$
one can get in two limiting cases $\hbar q \gg T$ and $\hbar q \ll T$,
where $q^2=\omega^2+c^2_s k^2$. In the first case the momentum
$q'\sim T/c_s\ll q$, and we can separate summation and integration over
$q'$ and put $q'=0$ in $G^{\Omega}(q-q')$ in (\ref{24}).
The result is extremely simple
\begin{eqnarray} \label{22}
\Pi_0(q)=4G^{\Omega}(q)=\frac{2{\cal J}(1+\gamma_{\bf k})}
{s(\omega^2+\omega^2_{0{\bf k}})}, \ \ \ q \gg k_T,\ \ \ k_T=T/c_s.
\end{eqnarray}
Notice, that it exceeds the quantum contribution in
(\ref{24}) $\Pi_0(q)=N/4q$ by the large parameter $16s{\cal J}/Nq$.
For small $q \ll c_s/a$ and $q \ll k_T$ our results coincide with \cite{Sac1}.

The dynamical spin susceptibility $\chi_{ij}(\omega,{\bf k})$ for
all values of $\omega$ and ${\bf k}$ can be calculated.
In the lowest order over $1/2s$ we can use the lowest order relation
\begin{equation}\label{23}
{\bf n}(\mbox{\boldmath{$\Omega$}}(\tau,l),{\bf M}(\tau,l),\tau,l)
\simeq e^{ia{\bf l}\cdot{\bf q}_{AF}}
\mbox{\boldmath{$\Omega$}}(\tau,l)-
[\mbox{\boldmath{$\Omega $}}(\tau,l)\times {\bf M}(\tau,l)],
\end{equation}
where ${\bf q}_{AF}=(\pi/a,\pi/a)$ is the AF vector (\ref{10}).
Calculating the average of two vectors ${\bf n}$ from (\ref{23}) we get
the dynamical spin susceptibility as a sum of two terms
$\chi_{ij}(\omega,{\bf k})=\delta_{ij}[\chi_A(\omega,{\bf k})+
\chi_F(\omega,{\bf k})]$.
The spin susceptibility $\chi_A(\omega,{\bf k})$ is responsible for the
AF fluctuations. It is proportional to the Green function
$G^{\Omega}_{q}$ analytically
continued and shifted by the AF vector ${\bf q}_{AF}$
\begin{eqnarray} \label{24}
\chi_A(\omega,{\bf k})=-\frac{Js^2z(1+\gamma_{{\bf k}^*})}
{2(\omega^2-\omega^2_{0{\bf k}^*}+i\omega\delta)},
\end{eqnarray}
where ${\bf k}^*={\bf k}-{\bf q}_{AF}$.
For the ferromagnetic spin susceptibility $\chi_F(\omega,{\bf k})$ we have
a loop expression which can be calculated on the basis of the thermal
fluctuation domination, as a result we have for $q \ge k_T$
\begin{eqnarray} \label{25}
\chi_F(\omega,{\bf k}) \simeq  -\frac{2s^2}{N}G^M(q)=
-\frac{Js^2z(1-\gamma_{\bf k})}
{N(\omega^2-\omega^2_{0{\bf k}}+i\omega\delta)}.
\end{eqnarray}

The theory of the spin fluctuations in the disordered QAF
at sufficiently low temperature $T \ll {\cal J}$
allows for the scale separation. In this case $k_T \ll \pi/a$
and the thermal fluctuations can be considered by the "renormalized
classical" manner \cite{Cha1}. The magnitude of the quantum fluctuations
at $q \le k_T$ is small in comparison with the classical fluctuations.
In this situation the parameters of the effective long wave, low frequency
sigma model are renormalized by the quantum fluctuations . This
renormalization is performed with respect to the parameter $1/2s$,
but the interaction of
the thermal fluctuations with the scales $|{\bf k}| \le k_T$ and
$\omega \le T$ is over parameter $1/N$, where $N$ is the number of
components of the ${\bf n}$ field of the long wave, low frequency
nonlinear sigma model.  This picture follows directly from the approach of
this paper.

 Unfortunately, the continuum approximation in time is not working when
we are calculating corrections to the basic approximation. The reason for
last observation lies in the canonical structure of the Lagrangian
(\ref{11}) and the Green function (\ref{16}): the sums over $\omega$
including this Green function are ambiguous and must be defined at the
final
time step $\Delta$. We only describe the basic approximation
and results shortly. They can be obtained on the basis of the SCS (\ref{1}).

Instead of the expression (\ref{7})
for the action $A({\bf n})$ we shall use more accurate expression
\begin{eqnarray} \label{26}
A({\bf n}) = -\sum_{j=0}^{N_{\tau}}\sum_{l}\Delta\left[{\cal L}_{kin}(j,l) +
{\cal H}(j,l)\right],
\end{eqnarray}
where $\tau=j\Delta$, and $\Delta N_{\tau}=\beta$.
Here, ${\cal L}_{kin}(j,l)$ consists of two parts
${\cal L}_{kin}={\cal L}_{mod}+{\cal L}_{pha}$. The
first term is pure real the second term is pure imaginary.

\begin{eqnarray} \label{27}
\Delta{\cal L}_{mod}=-s\ln\left[(1+\underline{\bf n}_a\cdot
{\bf n}_a)(1+\underline{\bf n}_b\cdot{\bf n}_b)/
4\right],
\end{eqnarray}
where ${\bf n}_p={\bf n}_p(j,l)$, $\underline{\bf n}_p={\bf n}_p(j+1,l)$,
for $p=a,b$.

The Lagrangian ${\cal L}_{pha}$ is not so simple
\begin{eqnarray} \label{28}
\Delta{\cal L}_{pha}=-\frac{s}{2}\sum_{p=a,b}
\ln\left(\frac{R_p\underline{R}_p^*}{R_p^*\underline{R}_p}\right),
\end{eqnarray}
where the quantity
$R_{p}=\underline{\bf n}_p\cdot{\bf m}_p+i\underline{\bf n}_p\cdot
{\bf k}_p$, where vectors ${\bf n},{\bf m},{\bf k}$ were defined at
the introduction of the SCS.

Expanding  ${\cal L}_{pha}$ in the vector ${\bf M}$ has
a rather complicated form but one can prove that it is regular
and contains only odd powers of ${\bf M}$.

The Hamiltonian ${\cal H}({\bf n})$ can be
obtained on the basis of the following relation
\begin{eqnarray} \label{29}
\mbox{\boldmath{${\cal  S}$}}(\underline{\bf n},{\bf n})=
\frac{<\underline{\bf n}|\hat{\bf S}|{\bf n}>}{<\underline{\bf n}
| {\bf n}>} = \frac{\underline{\bf n}+{\bf n}
-i[\underline{\bf n}\times{\bf n}]}{1+\underline{\bf n}\cdot{\bf n}},
\end{eqnarray}
for the matrix element of the spin operator $\hat{\bf S}$.
If we substitute them into the matrix element of the Heisenberg Hamiltonian
we obtain
\begin{eqnarray} \label{30}
{\cal H}({\bf n})=Js^2\sum_{l'\in <l>}
\mbox{\boldmath{${\cal  S}$}}(\underline{\bf n},{\bf n})
\cdot\mbox{\boldmath{${\cal  S}$}}
(\underline{\bf n}',{\bf n}'),
\end{eqnarray}

It is assumed that all vectors ${\bf n}_p, {\bf m}_a, {\bf k}_p$
for $p=a,b$ entering in Eqs. (\ref{27}-\ref{30})
are functions of the dynamical
variables $\mbox{\boldmath{$\Omega $}}$ and ${\bf M}$
according to Eq. (\ref{10}).

By expanding the Lagrangians ${\cal L}_{mod}$ (\ref{27}), ${\cal
L}_{pha}$ (\ref{28}), and the Hamiltonian (\ref{30}) in the vector
${\bf M}$ up to second order we get
\begin{eqnarray} \label{31}
&&\Delta{\cal L}_{kin}=s[1-\underline{\mbox{\boldmath{$\Omega$}}}\cdot
\mbox{\boldmath{$\Omega$}}+{\bf M}^2-\underline{\bf M}\cdot{\bf M}+
i(\underline{\mbox{\boldmath{$\Omega$}}}\cdot{\bf M}-
\mbox{\boldmath{$\Omega$}}\cdot\underline{\bf M})], \ \ \
\\ \nonumber
&&{\cal H}=Js^2\sum_{l'\in <l>}[\underline{\mbox{\boldmath{$\Omega$}}}\cdot
\mbox{\boldmath{$\Omega$}}-\mbox{\boldmath{$\Omega$}}\cdot
\mbox{\boldmath{$\Omega$}}'+\underline{\bf M}\cdot{\bf M}+
{\bf M}\cdot{\bf M}'-i(\underline{\mbox{\boldmath{$\Omega$}}}\cdot{\bf M}-
\mbox{\boldmath{$\Omega$}}\cdot\underline{\bf M})].
\end{eqnarray}

Now we present the result for the free energy $F_{QAF}$ of QAF which was
obtained in the lowest order in $1/2s$ on the basis
the Lagrangian (\ref{31}).
$F_{QAF}=F_{\Omega M}+F_{\lambda}$, and
\begin{eqnarray} \label{32}
&&F_{\Omega M}= -N_s{\cal J}+
2N_s\sum_{\bf k}\left\{\omega_{0{\bf k}}/2+
T\ln[1-\exp(-\omega_{0{\bf k}}/T)]\right\},
\\ \nonumber
&&F_{\lambda}=\frac{TN_s}{2}\sum_{\omega {\bf k}}
\ln\left[\frac{s(\omega^2+\omega^2_{0{\bf k}})\Pi_0(q)}
{2{\cal J}(1+\gamma_{\bf k})}\right].
\end{eqnarray}
Here $F_{\Omega M}$ is the free energy of the $\mbox{\boldmath{$\Omega$}}$ and ${\bf M}$ fields,
$F_{\lambda}$ is the free energy of the $\lambda$ field,
$2N_s$ is the number of the lattice sites,  and the polarization
operator $\Pi_0(q)$ is defined in (\ref{21}). The temperature dependent
part of the free energy (\ref{32}) at small temperatures $T\ll {\cal J}$
is proportional to $F_{AF}\approx N_sT^3/{\cal J}$.  Such contribution has
two origins: one from $F_{\Omega M}$ and another one from $F_{\lambda}$.

In the framework of the discretized time approach,
the first order corrections to the Green
functions of the $\mbox{\boldmath{$\Omega$}}$ and ${\bf M}$ fields were calculated and the
expression for the correlation length in the next order in $1/2s$
was obtained which  lies
completely in the framework of the concept of scales separation.
This results will be discussed in detail in a more complete publication.

\vskip 0.5cm  \noindent
{\Large \bf Acknowledgments} \\

We are grateful to A.V. Chubukov, I.V. Kolokolov, and S.Sachdev  for
stimulating discussions; C. Provid\^encia and V.R. Vieira for the accompany
discussions; A.M. Finkelstein, and P. Woelfle  for critical remarks.  One
of the authors (V.B.) is grateful for A.L. Chernyshev, L.V. Popovich, and
V.A. Shubin for the discussion and cooperation on an earlier step of this
work.
This work was supported in part by the Portuguese projects
PRAXIS/2/2.1/FIS/451/94, V.  B. was supported in part by the Portuguese
program PRAXIS XXI /BCC/ 11952 / 97, and in part by the Russian Foundation
for Fundamental Researches, Grant No 97-02-18546.

\end{document}